\documentclass{article}
\usepackage{ijcai26}

\usepackage{times}
\usepackage{soul}
\usepackage{url}
\usepackage[hidelinks]{hyperref}
\usepackage[utf8]{inputenc}
\usepackage[small]{caption}
\usepackage{graphicx}
\usepackage{amsmath}
\usepackage{amsthm}
\usepackage{booktabs}
\usepackage{algorithm}
\usepackage{algorithmic}
\usepackage[switch]{lineno}
\usepackage{multirow} % 加载 multirow 宏包实现单元格跨行
\usepackage{booktabs} % 可选，用于更美观的横线（可替换成 \hline）
\usepackage{makecell}
\usepackage{amssymb}
\usepackage{algorithm}
\usepackage{algorithmic}
\usepackage{amsmath}  % 用于数学符号
\title{ADCA: Attention-Driven Multi-Party Collusion Attack \\in Federated Self-Supervised Learning}

\author{
Jiayao Wang$^1$, Yiping Zhang$^1$, Jiale Zhang$^1$, Qilin Wu$^2$, Junwu Zhu$^{1}$ \thanks{Corresponding Author.}, and Dongfang Zhao$^3$\\
\affiliations
$^1$School of Information and Artificial Intelligence, Yangzhou University, China\\
$^2$School of Computing and Artificial Intelligence, Chaohu University, China\\
$^3$Tacoma School of Engineering and Technology, University of Washington, USA\\
\emails
\{wjiayao0203, jialezhang, Jwzhu\}@yzu.edu.cn,
mx120240592@stu.yzu.edu.cn,
lingqiw@126.com
dzhao@cs.washington.edu
}

\begin{document}

\maketitle

\begin{abstract}
Federated Self-Supervised Learning (FSSL) integrates the privacy advantages of distributed training with the capability of self-supervised learning to leverage unlabeled data, showing strong potential across applications. However, recent studies have shown that FSSL is also vulnerable to backdoor attacks. Existing attacks are limited by their trigger design, which typically employs a global, uniform trigger that is easily detected, gets diluted during aggregation, and lacks robustness in heterogeneous client environments. To address these challenges, we propose the Attention-Driven multi-party Collusion Attack (ADCA). During local pre-training, malicious clients decompose the global trigger to find optimal local patterns.  Subsequently, these malicious clients collude to form a malicious coalition and establish a collaborative optimization mechanism within it. In this mechanism, each submits its model updates, and an attention mechanism dynamically aggregates them to explore the best cooperative strategy. The resulting aggregated parameters serve as the initial state for the next round of training within the coalition, thereby effectively mitigating the dilution of backdoor information by benign updates. Experiments on multiple FSSL scenarios and four datasets show that ADCA significantly outperforms existing methods in Attack Success Rate (ASR) and persistence, proving its effectiveness and robustness.
\end{abstract}

\section{Introduction}
\begin{figure}
    \centering
    \includegraphics[width=0.8\linewidth]{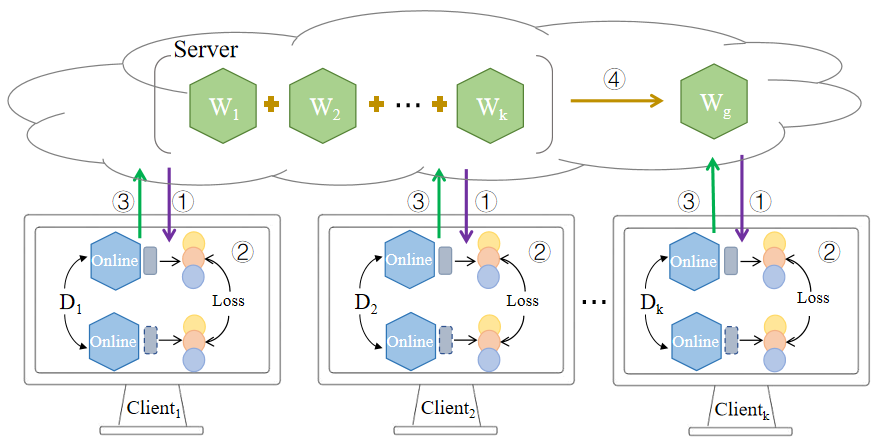}
    \caption{Federated Self-Supervised Learning Framework.}
    \label{fig:FSSL}
\end{figure}

The key idea of FSSL is to integrate the data privacy protection capabilities of Federated Learning (FL)~\cite{yang2019federated,dong2025smartfl} with the ability of Self-Supervised Learning (SSL)~\cite{caron2020unsupervised,he2020momentum} to learn from unlabeled data. By performing representation learning on unlabeled data locally across clients and aggregating model updates at the server, this approach not only effectively captures the intrinsic structure of data but also fundamentally avoids privacy leakage risks. Furthermore, due to its distributed nature, the method demonstrates robustness in data-heterogeneous scenarios, making it highly applicable to a wide range of privacy-sensitive tasks. Figure~\ref{fig:FSSL} presents an overview of the general FSSL framework, which incorporates an end-to-end training pipeline. The pipeline consists of the following steps: \textcircled{1} Model Broadcasting: At the beginning of communication round $t$, the server broadcasts the current global SSL model to the selected clients. \textcircled{2} Local Training: Selected client $k$ fine-tunes the received global model on its local dataset  $D_k$ to train a local model. \textcircled{3} Model Upload: Client $k$ uploads the trained local model to the server. \textcircled{4} Model Aggregation: The server aggregates all client models to generate the global model $W_g$. FSSL iterates through the above steps multiple times until the global model converges and achieves satisfactory performance.

While FSSL demonstrates significant value in privacy preservation and utilization of unlabeled data, its distributed collaborative training mechanism also introduces new security challenges, particularly the potential threat of backdoor attacks~\cite{zhang2021badfss,wu2024towards}. Most existing backdoor attacks against FSSL adhere to a centralized trigger design approach ~\cite{bagdasaryan2020backdoor,sun2019can}, where all malicious clients inject backdoors using a unified, global trigger pattern. However, such methods suffer from inherent limitations: on the one hand, the repeated implantation of a global trigger across models results in a highly consistent pattern, making the attack easily detectable at the model level; on the other hand, during the federated aggregation process, malicious updates are mixed with a large number of benign ones, causing the backdoor signal to be easily diluted, which severely compromises the persistence and effectiveness of the attack. In contrast, distributed triggers~\cite{liu2024beyond,feng2025sadba} decompose the trigger into multiple local patterns, each implanted by different clients. These patterns exhibit local heterogeneity, dynamic adaptability, and coupling with local data distributions, thereby enhancing stealth while improving resistance to aggregation-induced dilution. Nevertheless, research on the construction methods and attack efficacy of distributed triggers in FSSL remains underexplored and warrants further investigation.

To address these, we propose a new distributed backdoor attack method for FSSL, named the Attention-Driven multi-party Collusion Attack (ADCA). This approach introduces two core innovations: First, it designs a distributed injection mechanism based on trigger decomposition, which partitions the global trigger into multiple local patterns and disperses them across different clients, thereby training an encoder embedded with backdoor features through contrastive learning. Second, it proposes a local update strategy based on attention-driven interaction, where malicious clients dynamically integrate models from other attackers and the global model before each training round to update their initialization parameters, effectively mitigating the dilution of backdoor features during aggregation and significantly enhancing the persistence and stealthiness of the attack.

In summary, the main contributions of this paper are as follows:
\begin{itemize}

\item To our knowledge, this is the first work exploring the impact of trigger decomposition strategies on backdoor attack effectiveness in the FSSL scenario, addressing the research gap in this field regarding the association between distributed trigger pattern design and attack performance.

\item In the malicious coalition, we introduce an attention-driven interaction mechanism to dynamically fuse the global model with malicious client models, effectively mitigating the dilution of backdoor features during model aggregation.

\item We conduct systematic evaluations of ADCA on four benchmark datasets: CIFAR-10, STL-10, GTSRB, and CIFAR-100. The experimental results demonstrate that ADCA consistently outperforms existing methods under various settings and further exposes the limitations of current federated defense strategies in FSSL scenarios.
\end{itemize}

\begin{figure*}
    \centering
    \includegraphics[width=0.7\linewidth]{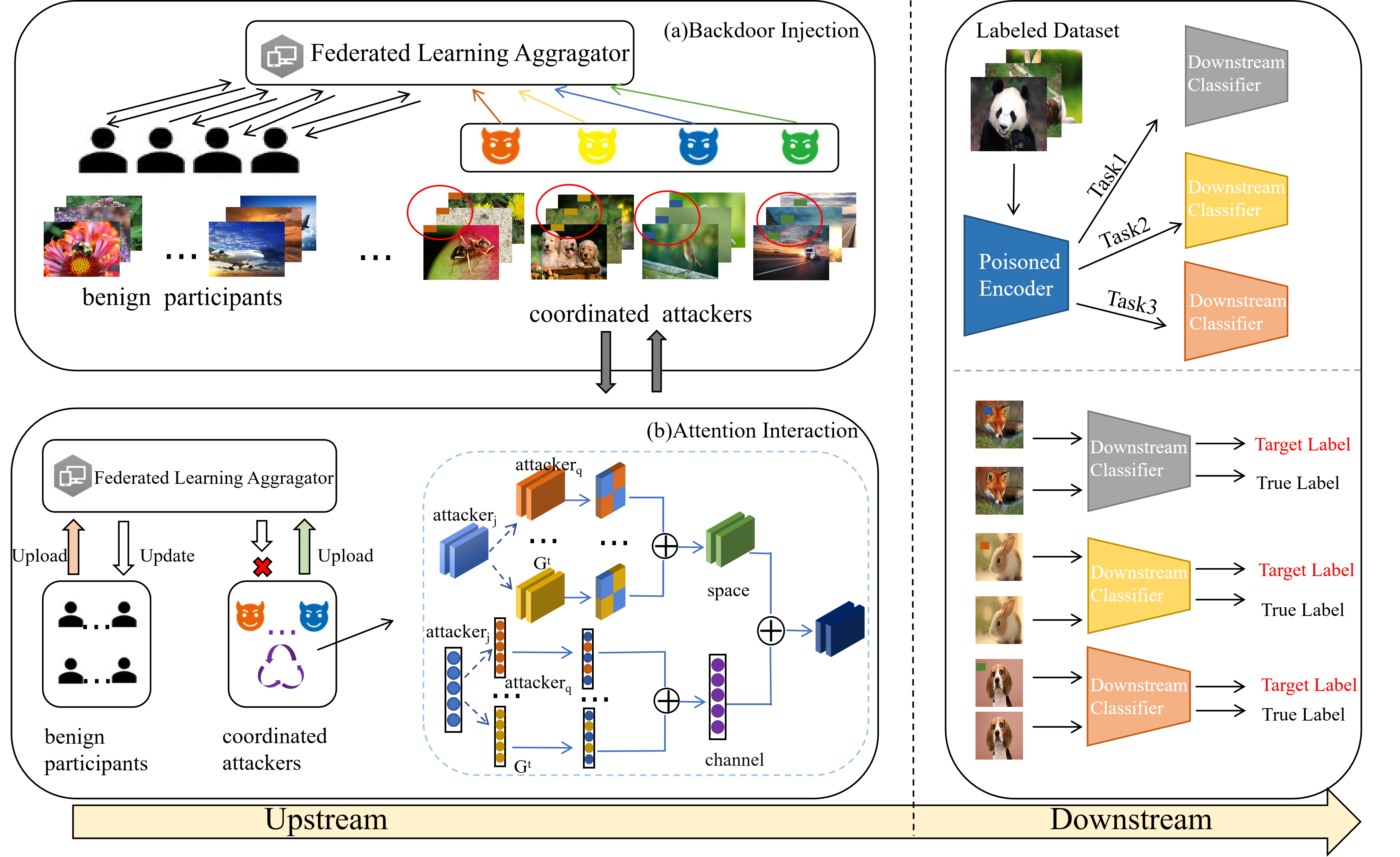}
    \caption{ADCA Overview.}
    \label{fig:ADCA}
\end{figure*}

\section{Related Work}
\subsection{Federated Self-Supervised Learning} 
FSSL has attracted growing attention in recent years due to its ability to collaboratively learn high-quality representations across distributed clients without requiring labeled data, while preserving data privacy. Early work~\cite{van2020towards} explored the feasibility of directly integrating classical SSL methods~\cite{grill2020bootstrap} into the federated learning framework. To address data heterogeneity, SSFL~\cite{he2021ssfl} and FedEMA~\cite{zhuang2022divergence} introduced personalized modeling strategies and momentum-based update mechanisms, respectively. For improving aggregation performance, FedCA~\cite{zhang2023federated} and L-DAWA~\cite{rehman2023dawa} proposed global dictionary learning and divergence-aware weighting schemes to enhance model fusion. In addition, encoder-specific aggregation~\cite{li2024resource}, knowledge distillation-based methods~\cite{kim2023protofl}, and dual-network architectures~\cite{li2023fedutn} have also been employed to improve communication efficiency and representation quality.
\subsection{Backdoor Attacks}
Backdoor attacks embed trigger patterns during training, causing models to behave normally on clean inputs but misclassify triggered inputs to attacker-defined targets. While traditional methods rely on labeled data to guide malicious behavior, such strategies are challenging in label-free SSL, motivating the development of tailored backdoor injection techniques. For example, Saha et al.~\cite{saha2022backdoor} embed triggers into target-class images, and apply crop-based augmentations for multi-view contrastive learning. BadEncoder~\cite{jia2022badencoder} fine-tunes pre-trained encoders using shadow datasets with embedded triggers to manipulate representations. Recent studies have also demonstrated the vulnerability of FSSL to backdoor attacks. BADFSS~\cite{zhang2021badfss} injects triggers via supervised contrastive learning and attention alignment to enable cross-client backdoor propagation. UBA~\cite{wu2024towards} employs intra-alliance aggregation to alleviate backdoor signal dilution during global model aggregation. However, most existing approaches rely on centralized triggers, which are prone to detection by manual inspection or automated defenses. In contrast, our proposed distributed trigger mechanism enhances both stealth and robustness while preserving high attack success rates.
\subsection{Backdoor Defenses}
Backdoor attacks are notoriously difficult to defend against due to their high stealth and potential for severe damage. To tackle this challenge, various defense strategies have been proposed. In FL, FLAME~\cite{nguyen2022flame} introduces an adaptive noise injection mechanism under the differential privacy framework, which adjusts the noise distribution to balance privacy preservation and model performance. FLTrust~\cite{cao2020fltrust} builds a reference model using trusted server-side data and assigns trust scores to local updates to guide aggregation. In SSL, PatchSearch~\cite{tejankar2023defending} detects abnormal image regions via local clustering to identify poisoned samples. PoisonCAM~\cite{qian2023erasing} clusters activation channels using clean data and suppresses or resets suspicious ones through masking to eliminate backdoor effects. In FSSL, EmInspector~\cite{qian2024eminspector} identifies malicious clients without requiring labels by leveraging embedding consistency between global and local views along with statistical characteristics. In this study, we systematically evaluate our proposed attack under these representative defense mechanisms.

\section{Threat Model}
\subsection{Attacker Objectives} 
Malicious clients aim to inject a backdoor that is effective, stealthy, and persistent. Effectiveness means that the backdoored model should classify inputs containing the trigger into the target class with high probability, i.e., achieving a high Attack Success Rate (ASR). Stealthiness requires that the model’s accuracy on the main task (ACC) does not degrade significantly, making the backdoor injection difficult to detect. Persistence indicates that the backdoor remains effective even after multiple rounds of training and aggregation.

\subsection{Attacker Capabilities} 
$M$ clients collude to launch a backdoor attack, forming a set of malicious clients denoted by  \( k_{\text{mal}} \). They have full control over their local models, training processes and datasets. However, they have no knowledge of the aggregation algorithm used by the server or the model updates from benign clients. If selected in a given training round, they download the broadcasted global model from the central server, and dynamically aggregate and learn the optimal collaboration pattern within the malicious alliance through an attention mechanism.

\section{Design of  ADCA}
Figure~\ref{fig:ADCA} illustrates our proposed ADCA framework, which adheres to the two-stage paradigm of FSSL: upstream pre-training and downstream prediction. Our work is concentrated on the upstream pre-training stage, whose architecture comprises both server and client components. On the server side, FedAvg~\cite{mcmahan2017communication} is employed to aggregate the uploaded model parameters; on the client side, we introduce our core contributions—namely, a distributed trigger injection strategy and an attention-driven collaborative optimization mechanism for malicious coalitions (The detailed algorithm is provided in Appendix F.).

\textbf{Step 1 (Distributed Trigger Injection): } Malicious clients construct distributed backdoor samples by decomposing a global trigger along three dimensions: trigger location(\( TL\)), trigger size(\( TS\)) and, trigger gap(\( TG\)), and inject the backdoor into the encoder via a contrastive learning approach.

\textbf{Step 2 (Attention-Driven Collaborative Optimization within a Malicious Alliance): } A cross-attention mechanism is employed within the malicious alliance to aggregate model updates from adversarial clients, and the aggregated parameters (rather than the global parameters) are used to initialize their models for the next round, preserving backdoor features during global aggregation.

Downstream Phase: The downstream dataset is fed into the pre-trained (potentially backdoored) model for classification testing.

In the following, we detail the two upstream components: Step 1 and Step 2.
\subsection{Distributed Trigger Injection}
\textbf{Motivation: Exploring Optimal Distributed Trigger Decomposition Strategies.} 
Centralized triggers~\cite{li2023embarrassingly,wang2024ghostencoder,dai2024federated} are susceptible to detection and aggregation-induced dilution in FSSL due to their globally uniform pattern. While distributed triggers can enhance stealth, the optimal strategy for decomposing them has not been systematically investigated. Although existing work~\cite{xie2019dba,qian2024eminspector} has demonstrated the feasibility of such triggers, it lacks a comprehensive investigation into how decomposition strategies affect attack performance and sustainability. To this end, this paper systematically explores the key impacts from three dimensions: trigger location (\( TL\)), size (\( TS\)), and gap (\( TG\)), as detailed in RQ2.
\subsubsection{Construction of distributed backdoor triggers.}
This paper proposes an innovative distributed trigger, building on DBA~\cite{xie2019dba}. Unlike traditional approaches, We decompose a single complete trigger into $M$ distinct sub-triggers, which are strategically embedded into the datasets of different clients, thereby achieving a more covert and significantly efficient attack effect. This process is precisely defined by the following transformation function:

\begin{equation}
R(x, e, \phi^*) = x \otimes (1 - T_{\phi^*}) + e \otimes T_{\phi^*},
\end{equation}
where \( x \) denotes the original image, \( e \) represents the trigger pattern, and \( \phi^* \) is the injection strategy parameter, including \( TL\), \( TS\), and \( TG\). The mask matrix \( T_{\phi^*} \), generated from \( \phi^* \), is a binary matrix used to indicate the trigger region. The symbol \( \otimes \) denotes Hadamard (element-wise) multiplication. Specifically, the function \( R(\cdot) \) injects the trigger at positions where the mask equals 1 and retains the original pixels where the mask equals 0. As a result, the output is a backdoor sample carrying localized trigger signals, used for attack training.

\begin{figure}
    \centering
    \includegraphics[width=1\linewidth]{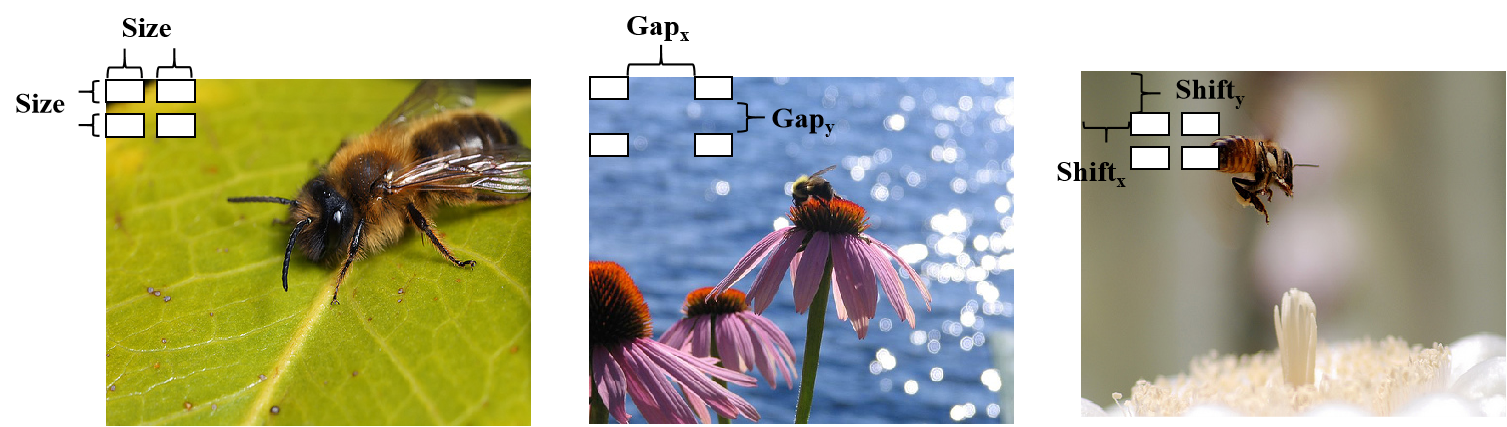}
    \caption{The trigger factors (size, gap, and position) in the backdoor image.}
    \label{fig:Triger}
\end{figure} 

Figure~\ref{fig:Triger} illustrates three key spatial attributes of image-based triggers: \( TL\), \( TS\), and \( TG\). For unified modeling and visualization, all local triggers in the figure are set to have the same rectangular shape, and four sub-triggers are uniformly overlaid on a single image solely to aid in explaining the semantic meaning of each dimension. This figure demonstrates the constraint relationships among the trigger injection strategy parameters, which are mathematically formulated as follows:

\begin{equation}
\phi^{*}=\begin{cases} 
TL \in [0, H] \times [0, W] \\
TS \in (0, \min(H, W)] \\ 
TG \leq \sqrt{H^{2} + W^{2}}
\end{cases},
\end{equation}

\noindent
where H and W represent the height and width of the image, respectively, \( TL \) allows the trigger to be placed anywhere in the image, \( TS \) controls the size of the trigger region, with its upper bound being the minimum dimension of the image. \( TG \) defines the minimum distance between triggers, with its upper bound being the diagonal length of the image.

\subsubsection{Optimization of the local backdoor encoder.}
Based on BadEncoder, during the local training phase on malicious clients, we utilize a shadow dataset composed of both clean samples, and backdoor samples embedded with distributed triggers to train the backdoor encoder under a robust contrastive learning framework (detailed descriptions are provided in Appendix A), this process aims to jointly optimize two key, interrelated objectives.

\noindent
\textbf{Backdoor Attack Effectiveness Objective.} To enhance attack effectiveness, the attacker aims to optimize the local model such that it accurately identifies trigger-embedded inputs, and maps them to a predefined target class, without compromising performance on the primary task. This objective can be formally defined by the following loss function:

\begin{equation}
\begin{split}
L_{\text{eff}} &= -E_{x\sim P_{\text{adv}}} \Bigl[ S\bigl(f(x \oplus e_k, \widetilde{\theta}_k), f(x_t, \widetilde{\theta}_k)\bigr) \\
&\quad + \beta \cdot S\bigl(f(x_t, \widetilde{\theta}_k), f(x_t, \theta_g)\bigr) \Bigr],
\end{split}
\end{equation}
where \( S(\cdot, \cdot) \) is the cosine similarity function and \( f(\cdot, \cdot) \) denotes the model’s feature representation. \(P_{\text{adv}} \) denotes the joint distribution of backdoor samples, and \( x \) denotes the clean input. \( e_k \) is the distributed trigger for client \( k \), derived from the decomposition strategy. \(\tilde{\theta}_k\) represents the model parameters for the malicious client \(k\). \( \theta_g \) is the global model at the current round, and \( \beta = 0.8 \) is a balancing factor for consistency.

\noindent
\textbf{Feature Stealthiness Objective.} To enhance stealthiness, malicious clients aim to align the representations of clean inputs generated by the backdoored model with those produced by the benign global model. This objective can be formalized as:
\begin{equation}
\begin{split}
L_{\text{stealth}} &= -E_{(x, x^+) \sim P_{\text{clean}}} \Bigl[ S\bigl(f(x, \widetilde{\theta}_k), f(x, \theta_g)\bigr) \\
&\quad + S\bigl(f(x^+, \widetilde{\theta}_k), f(x^+, \theta_g)\bigr) \Bigr],
\end{split}
\end{equation}
where \( x^+ \) its augmented view, and\( P_{\text{clean}} \) denotes the joint distribution of clean sample pairs.

\noindent
\textbf{Joint Optimization Objective.} To balance attack effectiveness and feature stealthiness, we introduce weighting coefficients \( \lambda_1 \) and \( \lambda_2 \) to coordinate the sub-objectives. The overall optimization objective is defined as:

\begin{equation}
\min_{\tilde{\theta}_k} L = \lambda_1 \cdot L_{\mathrm{eff}} + \lambda_2 \cdot L_{\mathrm{stealth}}.
\end{equation}

\subsection{Attention-Driven Collaborative Optimization within a Malicious Alliance}
\textbf{Motivation: Backdoor Persistence and Dilution in FSSL.}
In FSSL settings, prior works~\cite{bagdasaryan2020backdoor,xie2019dba} typically require malicious clients to synchronize their local models with the current global model at the beginning of each round. However, this strategy may dilute backdoor signals during aggregation with benign updates, reducing their effectiveness. Formally:
\begin{equation}
\left\| \frac{ \sum_{a \in k_{\text{mal}}} \Delta_a + \sum_{b \in k_{\text{benign}}} \Delta_b }{|n|} \right\|_{\text{backdoor}} \ll \left\| \Delta_{\text{mal}} \right\|_{\text{backdoor}},
\end{equation}
where $n$ denotes the total number of clients selected for training in the current communication round, $k_{\text{mal}}$ and $k_{\text{benign}}$ represent the sets of malicious and benign clients, respectively.

In addition, recent defenses detect abnormal embedding distributions by measuring the divergence between backdoor and clean samples in the global model. The detection metric is defined as:
\begin{equation}
\text{PCM} = \left\| \frac{1}{n} \sum_{i=1}^{n} z_i^{\text{backdoor}} - \frac{1}{n} \sum_{i=1}^{n} z_i^{\text{clean}} \right\|,
\end{equation}
where $z_i^{\text{backdoor}}$ and $z_i^{\text{clean}}$ denote the feature embeddings of backdoor and clean samples, respectively.

To address these challenges, UBA enhances the persistence of backdoor updates in the global model via intra-alliance model aggregation, but its reliance on simple weighted averaging is limited. Inspired by MSCA~\cite{xiang2025federated}, we propose an Attention-Driven Collusion Attack (ADCA) that fosters feature synergy among malicious clients through spatial and channel attention interactions. Specifically, ADCA aggregates current-round malicious models to initialize local models for attack consistency, while incorporating the global model as a constraint to avoid exposure from distribution deviation. This approach boosts backdoor expressiveness, counters aggregation dilution and high-dimensional defense detection, and maintains stealthiness. Performance improvements are demonstrated in experiment RQ1. The detailed aggregation method is as follows.

\begin{table*}[ht] 
\renewcommand{\arraystretch}{1.5}  % 调整行高
\centering 
\setlength{\tabcolsep}{2.7pt} % 更小的列间距
\begin{tabular}{ccccccccccccccc} 
\toprule
\multirow{2}{*}{\makecell{Pre-train\\Dataset}} & 
\multirow{2}{*}{\makecell{Downstream\\Dataset}} & 
\textbf{Benign} &
\multicolumn{2}{c}{\textbf{BadEncoder}} & 
\multicolumn{2}{c}{\textbf{BADFSS}} & 
\multicolumn{2}{c}{\textbf{DBA}} & 
\multicolumn{2}{c}{\textbf{FCBA}} & 
\multicolumn{2}{c}{\textbf{UBA}} & 
\multicolumn{2}{c}{\textbf{ADCA}} \\
\cmidrule(lr){3-15} 
& & \textbf{ACC} & \textbf{ACC} & \textbf{ASR} & \textbf{ACC} & \textbf{ASR} & \textbf{ACC} & \textbf{ASR} & \textbf{ACC} & \textbf{ASR} & \textbf{ACC} & \textbf{ASR} & \textbf{ACC} & \textbf{ASR} \\ 
\midrule 
\multirow{2}{*}{CIFAR-10} & STL-10 & 76.91 & 74.29 & 64.37 & 74.51 & 72.46 & 73.76 & 63.78 & 71.43 & 66.32 & 72.37 & 70.85 & 75.24 & 96.57 \\  
& GTSRB & 81.13 & 78.45 & 62.78 & 75.53 & 61.33 & 76.54 & 57.98 & 73.15 & 75.13 & 70.86 & 65.33 & 80.56 & 96.11 \\ 
\midrule 
\multirow{2}{*}{STL-10} & CIFAR-10 & 83.64 & 77.34 & 70.95 & 76.64 & 61.75 & 78.43 & 65.32 & 80.52 & 63.27 & 79.56 & 69.14 & 82.78 & 94.93 \\  
& GTSRB & 73.78 & 69.28 & 65.94 & 70.37 & 60.28 & 68.67 & 56.34 & 70.34 & 64.25 & 72.43 & 61.72 & 73.69 & 85.56 \\ 
\midrule 
\multirow{2}{*}{CIFAR-100} & CIFAR-10 & 71.51 & 70.38 & 65.96 & 70.23 & 63.18 & 70.14 & 64.56 & 67.32 & 59.56 & 68.21 & 63.25 & 71.04 & 95.27 \\  
& GTSRB & 75.67 & 72.93 & 61.53 & 66.39 & 62.57 & 73.87 & 60.52 & 73.28 & 61.42 & 74.65 & 66.43 & 74.53 & 91.78 \\ 
\bottomrule 
\end{tabular}  
\caption{Attack Performance Comparison on Different Datasets.} 
\label{tab:performance} 
\end{table*}

By analyzing the characteristics of client parameters in spatial distribution and channel activation, ADCA assigns dynamic weights to each malicious client. There are $M$ malicious clients, and the parameters of the \( j \)-th malicious client are updated as \( \Delta \theta_j \). The parameter updates are performed through convolutional kernel weight reweighting, which proceeds through layer pooling to obtain the malicious parameters' distribution in the spatial dimension. Specifically, for convolutional kernel reweighting, the weight tensor \( W \in R^{C_{\text{out}} \times C_{\text{in}} \times K_h \times K_w} \) performs self-optimization on the spatial dimension. It generates spatial feature vectors \( S_j \in R^{ K_h \times K_w} \) to represent the region of interest activated by the backdoor. Using the spatial dimension's uniform pooling, the statistical features \( F_j \in R^{D \times H \times W} \) are extracted. These features are then used to generate the channel features \( C_j \in  R^D \), which indicate the activation strength of the specific trigger after the backdoor.

\noindent
\textbf{Spatial interaction score.}  
For the \( j \)-th malicious client, the similarity between its spatial feature vector \( S_j \) and the spatial feature vectors of the remaining malicious clients and the previous round's global model is calculated. The result is normalized using the sigmoid activation function applied to the scaled inner product, defined as follows:

\begin{equation}
\alpha_j^{\text{spatial}} = \frac{1}{M} \sum_{p \neq j} \text{sigmoid}\left( \frac{S_j \cdot S_p^T}{\sqrt{d}} \right),
\end{equation}
where \( S_p \) represents the spatial feature vectors derived from other malicious clients and the global model of the last communication round, and \( d \) is the scaling factor.

\noindent
\textbf{Channel interaction score.}  
To capture the correlation between different clients at the channel feature level, the following channel interaction attention score is defined:
\begin{equation}
\alpha_j^{\text{channel}} = \frac{1}{M} \sum_{p \neq j} \text{sigmoid}\left(\text{MLP}(C_j \oplus C_p)\right),
\end{equation}
where \( C_j \) and \( C_p \) denote the channel feature vectors of client\( j \), other malicious clients and the global model from the previous round, respectively. \( \oplus \) denotes the vector concatenation operation, and MLP is a multilayer perceptron used to extract the channel-level interaction representation.

By summing the spatial interaction score and the channel interaction score, the aggregation weight for malicious client \( j \) is obtained:

\begin{equation}
\alpha_j = \alpha_j^{\text{spatial}} + \alpha_j^{\text{channel}}.
\end{equation}

Based on this, the initialization model update formula for malicious clients in the next round is defined as follows:

\begin{equation}
\tilde{\theta}^{t+1} = \tilde{\theta}^t + \sum_{j=1}^{M} \frac{\alpha_j}{\sum_{k=1}^{M} \alpha_k} \cdot \Delta \theta_j.
\end{equation}

\noindent
\section{Evaluation}
To demonstrate the effectiveness, stability, and robustness of our method, we implemented ADCA using PyTorch and compared its performance against state-of-the-art backdoor attack baselines. All experiments were conducted on an NVIDIA 4090 GPU. We designed comprehensive experiments to address the following four research questions:

\noindent
\textbf{RQ1 (Effectiveness of ADCA):} Can ADCA successfully inject a backdoor into FSSL?

\noindent
\textbf{RQ2 (Trigger Factors Impact of ADCA):} How do ASR and ACC change under different trigger factors in ADCA?

\noindent
\textbf{RQ3 (Stability of ADCA):} Can ADCA maintain stable performance under different settings?

\noindent
\textbf{RQ4 (Robustness of ADCA):} Can ADCA effectively resist existing defense methods?

\subsection{Experimental Setup}
\textbf{Datasets and Federated Environment.} Four datasets are employed in the experiments including CIFAR-10~\cite{krizhevsky2009learning}, CIFAR-100~\cite{krizhevsky2009learning}, STL-10~\cite{coates2011analysis} and GTSRB~\cite{stallkamp2012man}. More details about the used datasets can be found in Appendix B. For the IID setting, each client contains an equal amount of samples from each class. For the Non-IID setting~\cite{zhu2021data}, we follow prior art that models Non-IID data distributions using a Dirichlet distribution Dir(\( \alpha \)), where a smaller \( \alpha \) indicates higher data heterogeneity.

\noindent
\textbf{Evaluation Metrics.} We use ACC and ASR to evaluate our ADCA. ACC represents the classification accuracy of  global self-supervised model on clean samples, ASR indicates the proportion of samples with triggers that are misclassified as the target class specified by the attacker. A successful backdoor attack should maximize ASR while maintaining a high ACC, thereby achieving covert contamination of the model.

\noindent
\textbf{Baselines.} To comprehensively evaluate attack performance in the FSSL scenario, we select two distributed backdoor attack methods (DBA, FCBA~\cite{liu2024beyond}) originally designed for FL, and migrate them to the SSL framework for evaluation. We also include UBA—a distributed backdoor attack method tailored for FSSL—as a comparative baseline. For fairness, all attacks adopt a unified global trigger: split into equal local trigger blocks, injected by malicious clients to launch collaborative attacks. Additionally, to compare distributed and centralized trigger performance, we include BadEncoder and BADFSS in our experiments.

\noindent
\textbf{Implementation Details.} We use SimCLR for SSL and ResNet-18~\cite{he2016deep} as encoder backbone. Experiments are on an FSSL system of 25 clients (20 benign, 5 malicious). The server randomly selects 10 clients per communication round for local training and global aggregation, causing variant malicious client participation. Following DBA, the global trigger is split into local patterns for injection by malicious clients. Distributed triggers are used in training (each malicious client injects its specific pattern part), and the full global trigger is applied in testing to measure attack success rates. All experiments set  \( \lambda_1 \), \( \lambda_2 \)=1 in Equation (5), adopt SGD (lr=0.05) for attacks, and run 25 communication rounds with 3 local training epochs per round for clients.

\begin{figure}
    \centering
    \includegraphics[width=0.8\linewidth]{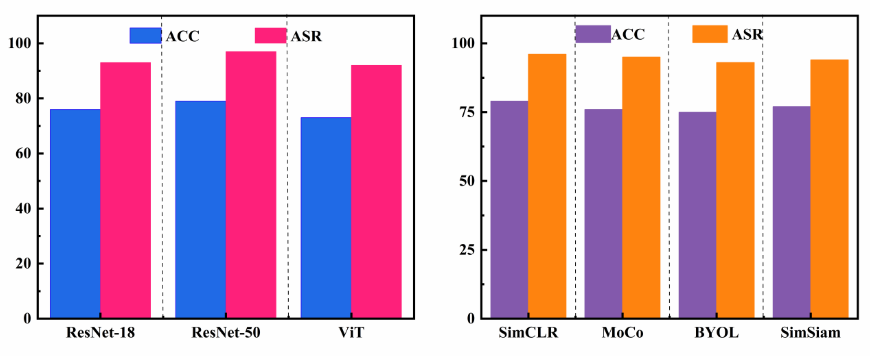}
    \caption{Experimental results for different encoder architectures and SSL algorithms.}
    \label{fig:different}
\end{figure}

\subsection{Evaluation of Effectiveness (RQ1)}
\textbf{Comparison with Baseline Attack Methods.} Table~\ref{tab:performance} compares ADCA with five baseline backdoor methods under standard SSL settings with different pre-training and downstream datasets. Using CIFAR-10 for pre-training and STL-10 downstream, ADCA achieves the highest ASR (96.57\%), surpassing the weakest baseline DBA (56.34\% ASR) by 40.23\%. Moreover, distributed triggers outperform centralized triggers such as BadEncoder and BADFSS. As dataset scale increases, ADCA maintains a consistently high ASR, demonstrating more robust and superior performance across various pre-training and downstream dataset combinations compared to other backdoor attacks.

\begin{figure}
    \centering
    \includegraphics[width=0.9\linewidth]{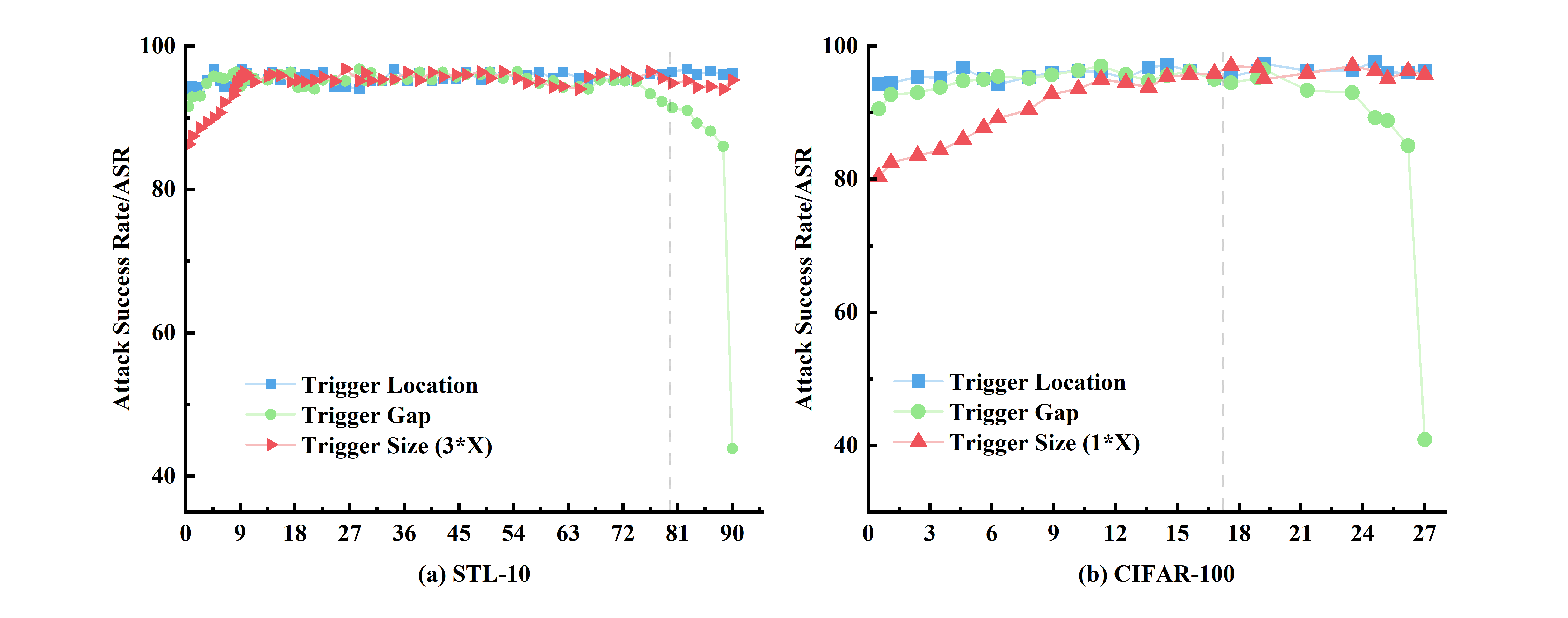}
    \caption{Experimental results on STL-10 and CIFAR-100, showing the impact of different triggers on ASR. Results on CIFAR-10 are detailed in Appendix C.}
    \label{fig:picture}
\end{figure}

\noindent
\textbf{Effectiveness Across Different SSL Algorithms.} To evaluate the generalizability of ADCA, we conduct experiments in a FSSL setting under four representative SSL methods: SimCLR~\cite{chen2020simple}, MoCo~\cite{he2020momentum}, BYOL~\cite{grill2020bootstrap}, and SimSiam~\cite{chen2021exploring}. As shown in Figure~\ref{fig:different}, ADCA achieves consistently high attack success rates across all SSL methods, demonstrating its robustness, adaptability and strong generalization ability.

\noindent
\textbf{Effectiveness Across Different Encoder Architectures.} To evaluate ADCA across different encoder architectures, we conduct experiments on STL-10 using ResNet-18~\cite{he2016deep}, ResNet-50~\cite{he2016deep}, and ViT~\cite{dosovitskiy2020image}. As shown in Figure \ref{fig:different}, ADCA consistently achieves successful backdoor injection while preserving high classification accuracy, confirming its strong adaptability and generalization across diverse backbones.

\subsection{Trigger Factor Evaluation (RQ2)}
\textbf{Impact of Trigger Position.} Based on DAB, we shift the global trigger from the top-left corner of the image to the center and then to the bottom-right corner. In Figure~\ref{fig:picture}, the dashed line indicates that the trigger has reached the image boundary and starts to move along the right edge. Experimental results show that \( TL \) has a relatively minor impact on ASR. This can be attributed to the fact that the three datasets contain color images with rich textures, whose high structural complexity and semantic diversity allow the backdoor trigger to blend well into the image at different locations, thereby enhancing robustness against positional variation.

\noindent
\textbf{Impact of Trigger Gap.} For the four local trigger patterns located at the four corners of the image, which correspond to the maximum trigger gap as shown in Figure~\ref{fig:picture}, ASR decreases significantly. This performance degradation occurs because the local receptive field of convolutional neural networks cannot simultaneously capture multiple spatially distant triggers. Consequently, the model fails to aggregate these dispersed features into a consistent global representation during contrastive learning or other self-supervised learning tasks.

\noindent
\textbf{Impact of Trigger Size.} Experimental results show that, in image datasets, larger trigger sizes tend to yield higher ASR. However, once \( TS \) becomes sufficiently large, the ASR begins to stabilize, indicating that excessively large triggers provide little to no additional benefit.

In summary, while the ASR is most robust to changes in \( TL \) and most sensitive to \( TG \), the gains exhibit saturation beyond a certain \( TS \).

\subsection{Stability Evaluation (RQ3)}
\textbf{Stability under Different Data Distribution Settings.} In FL scenarios, diverse Non-Independent and Identically Distributed (Non-IID) data distributions constitute a crucial and realistic setup. According to our experimental design, we employ Dir(\( \alpha \)) to simulate Non-IID data distributions, where \( \alpha \) varies across 0.05, 0.1, and 10. Specific experiments are provided in Appendix D, with results showing ADCA retains stable performance in these heterogeneous configurations. 
\begin{figure}
    \centering
    \includegraphics[width=0.45\linewidth]{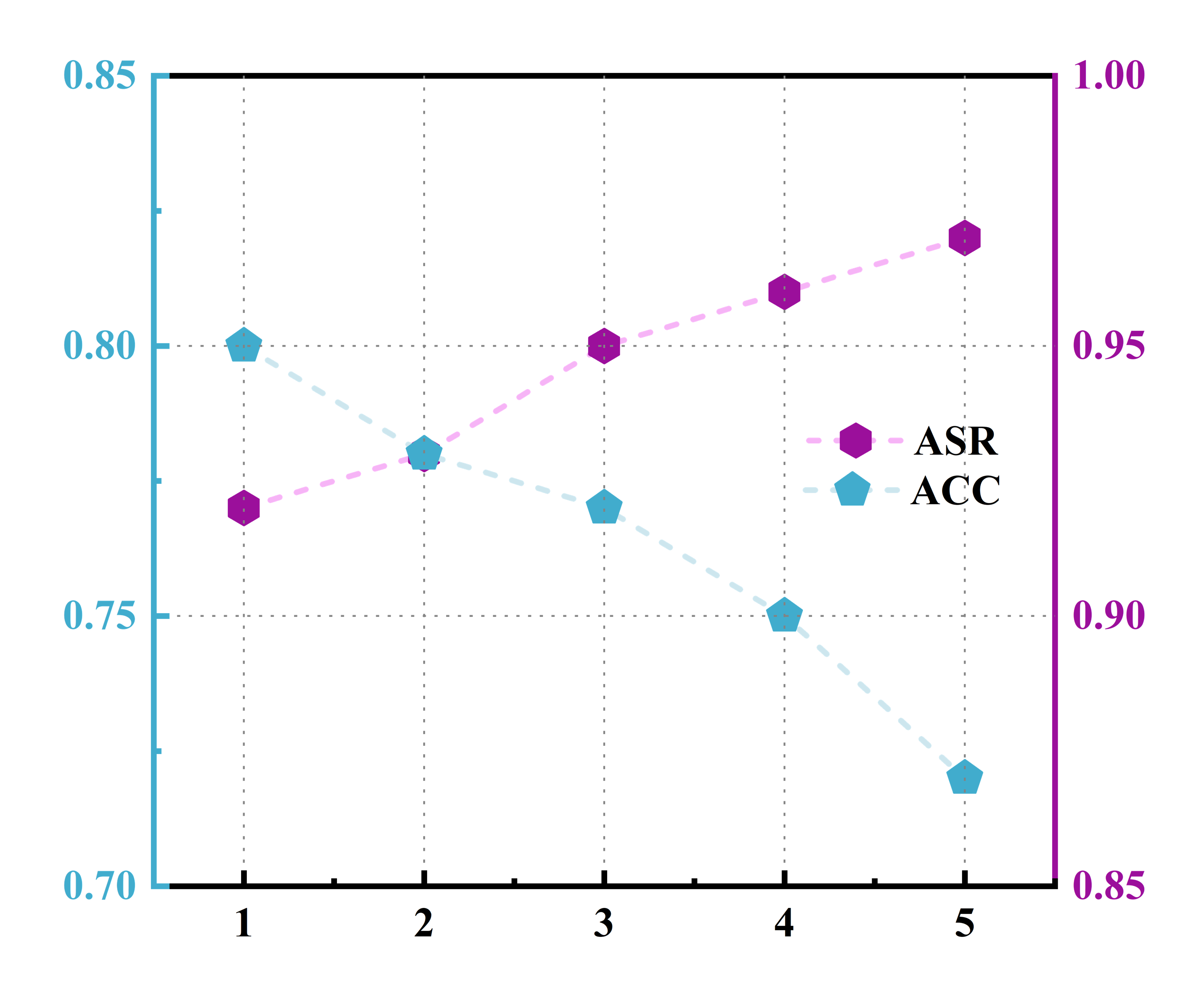}
    \caption{Performance of ADCA with different ratios of malicious clients.}
    \label{fig:fl}
\end{figure}

\noindent
\textbf{Stability under Different Numbers of Clients.} The number of clients is another critical setup in FL scenarios. To verify the applicability of our proposed method, we evaluate the impact of varying client numbers on ADCA's performance across three datasets. Specific experimental details are presented in Appendix E, and the results demonstrate that ADCA maintains a high ASR across all settings on the CIFAR-10, STL-10, and CIFAR-100 datasets, with only a slight drop in ACC.

\noindent
\textbf{Impact of Different Malicious Client Ratios on Stability.} According to our FL setup, 10 clients were deployed (each with equal samples), including 5 malicious ones; at least one malicious client was guaranteed to be selected for training per round. Figure~\ref{fig:fl} shows the variation trends of ASR and ACC under this setup. Results indicate that ADCA’s attack performance rises with the malicious client ratio—specifically, it performs better at higher malicious client proportions.

\subsection{Robustness Evaluation (RQ4)}
Table~\ref{defence} presents the performance of ADCA under five mainstream defense methods: FLAME, FLTrust, PatchSearch, PoisonCAM, and EmInspector. Experimental results show that although these defense strategies offer a certain degree of robustness in the FSSL setting, they are still insufficient to effectively mitigate the impact of ADCA. Specifically, on the three benchmark datasets of CIFAR-10, STL-10, and CIFAR-100, 
most defense methods (e.g., FLAME and FLTrust) exhibit a slight decrease in ACC while their ASR remain at a high level ranging from 85\% to 93\%. This phenomenon indicates that distributed backdoors have been successfully embedded into the global model through the utilization of attention mechanisms.

These findings suggest that ADCA retains strong robustness and stealth under existing defense schemes. This highlights the pressing need for developing more targeted defense strategies to counter distributed backdoor threats in FSSL.

\begin{table}[ht]
\centering
\setlength{\tabcolsep}{3.7pt} % 更小的列间距
\begin{tabular}{@{}lcccccc@{}} % 根据实际列数和内容调整列格式，这里示例列格式仅供参考
\hline
\multirow{2}{*}{\centering {\makecell{Defence\\Method}}} & \multicolumn{2}{c}{CIFAR-10} & \multicolumn{2}{c}{STL-10} & \multicolumn{2}{c}{CIFAR-100} \\
\cline{2-7}
 & ACC & ASR & ACC & ASR & ACC & ASR \\
\hline
FLAME & 66.45 & 92.76 & 65.32 & 91.25 & 61.42 & 87.53 \\
FLTrust & 65.26 & 93.14 & 64.15 & 92.36 & 60.37 & 89.65 \\
PatchSearch & 63.72 & 90.45 & 64.63 & 91.06 & 58.13 & 88.34 \\
PoisonCAM & 64.43 & 89.73 & 63.86 & 90.23 & 57.65 & 85.23 \\
EmInspector & 67.73 & 51.36 & 66.95 & 50.43 & 60.16 & 45.36 \\
\hline
\end{tabular}
\caption{Performance of ADCA under different defense methods.}
\label{defence}
\end{table}

\subsection{Ablation Study}
This section conducts an ablation study on the key components of ADCA to assess their contributions to the attack performance. As shown in Table~\ref{tab:ablation}, removing any single module leads to a significant drop in ASR on the downstream task. Specifically, when Distributed Trigger and Attention Interaction are removed individually, the ASR of ADCA drops from 96.34\% to 69.58\% and 87.16\%, respectively. This trend aligns with our previous observations, further confirming the critical role these two modules play in ensuring attack effectiveness. These results clearly demonstrate that each core component is essential for the success of the backdoor injection strategy.

\begin{table}[ht]
\centering
\setlength{\tabcolsep}{3.7pt}
\begin{tabular}{cccc}
\hline
\textbf{Distributed  Trigger} & \textbf{Attention Interaction} & \textbf{ACC} & \textbf{ASR} \\
\hline
$\checkmark$  & $\checkmark$ & 69.24 & 96.34 \\ \hline
$\checkmark$ &             & 72.13 & 69.58 \\ \hline
           &$\checkmark$ & 70.43 & 87.16 \\
\hline
\end{tabular}
\caption{Performance of Ablation Studies.}
\label{tab:ablation}
\end{table}

\section{Conclusion}
This paper proposes ADCA, a novel distributed backdoor attack for FSSL. By designing localized triggers and an attention-driven interaction mechanism, ADCA effectively tackles critical challenges of backdoor dilution and functional degradation in existing attacks. Experiments show that ADCA significantly outperforms baselines across multiple settings, demonstrating superior attack success rates and stability. Furthermore, we show that existing defenses are ineffective against ADCA, underscoring the need for tailored strategies to secure FSSL.

\section*{Ethical Statement}
The research presented in this paper investigates security vulnerabilities in FSSL systems. Our goal is to enhance system robustness by demonstrating a new attack vector (ADCA). We clarify that this work is intended solely for academic and defensive purposes. All experiments were performed in a simulated environment using public datasets (e.g., CIFAR-10, GTSRB), adhering to standard ethical guidelines. No real-world systems were compromised, and no human participants were involved.

\section*{Acknowledgements}
This work was supported by grants 24KJB520042 (Jiangsu), 2025YSZ-017 (Yangzhou), 2023SGJ014 (Hefei), COGOS2023HE01 (iFLYTEK), Y202352288 (Zhejiang), and 2023AY11057 (Jiaxing), as well as by resources from Microsoft Azure and the NSF-supported Chameleon testbed. Additionally, it was supported by the Anhui Provincial University Outstanding Research and Innovation Team Program (No.2024AH010022).

\bibliographystyle{named}
\bibliography{ijcai26}

\clearpage
\appendix
\section{Overview of Self-Supervised Contrastive Learning and Its Backdoor Attack Framework}
Self-supervised contrastive learning is a method that learns effective representations from data without manual annotation. Taking the SimCLR framework as an example, it generates two correlated views of the same sample through random augmentations as a positive pair. These views are then mapped into a representation space via an encoder and a projection head. The training process optimizes a contrastive loss to pull the representations of positive pairs closer together while pushing away those of other samples, enabling the model to learn transferable general-purpose representations. This paradigm is commonly used as a foundational pretraining approach in federated self-supervised learning.

\begin{figure}
    \centering
    \includegraphics[width=0.7\linewidth]{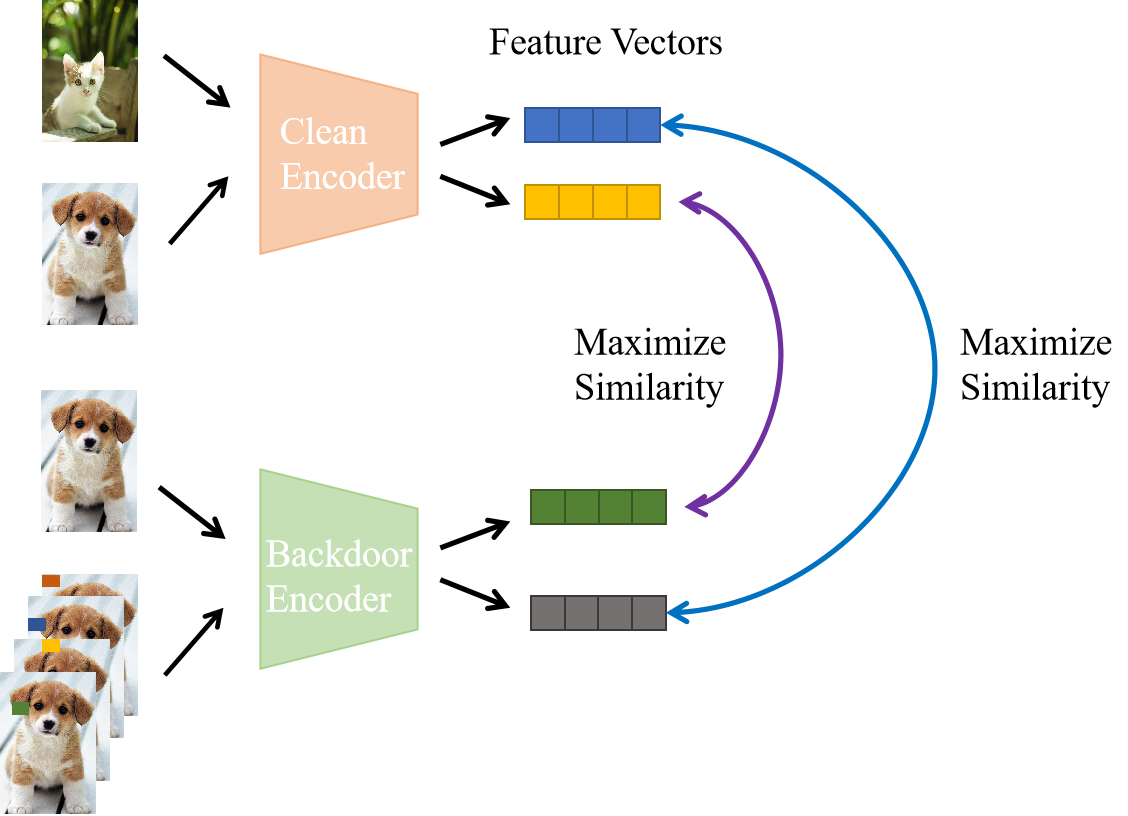}
    \caption{Backdoor Training Framework for SSL.}
    \label{fig:f2}
\end{figure}

However, this framework can be exploited by malicious participants to implant backdoors. As illustrated in Figure~\ref{fig:f2}, in a backdoor-injected contrastive learning framework, the attacker embeds a trigger during the local data preprocessing stage to construct a “backdoor-positive pair” consisting of a triggered view and a clean view. During training, the attacker adjusts the optimization direction of the loss so that the representation of the triggered view is pulled closer to the target class cluster while maintaining consistency with the representation of the clean view. This process is performed locally by malicious clients. Subsequently, through parameter uploading and global aggregation, the hidden backdoor behavior is continuously infused into the federated model, achieving a stealthy attack without compromising the primary task performance.

\section{Detailed Dataset Information}
We employ the following datasets for method evaluation:\\
    \textbf{CIFAR-10:} 10 classes, 50,000 training and 10,000 test images. A balanced benchmark for initial validation.\\
    \textbf{CIFAR-100:} 100 classes, 50,000 training and 10,000 test images. Introduces fine-grained classification challenges.\\
    \textbf{GTSRB:} 43 classes, 39,200 training and 12,600 test images. Represents a practical, real-world traffic sign recognition task.\\
    \textbf{STL-10:} 10 classes, 5,000 training and 8,000 test images. A standard benchmark for unsupervised and self-supervised learning.

\section{Trigger Factor Evaluation}
\begin{figure}
    \centering
    \includegraphics[width=0.7\linewidth]{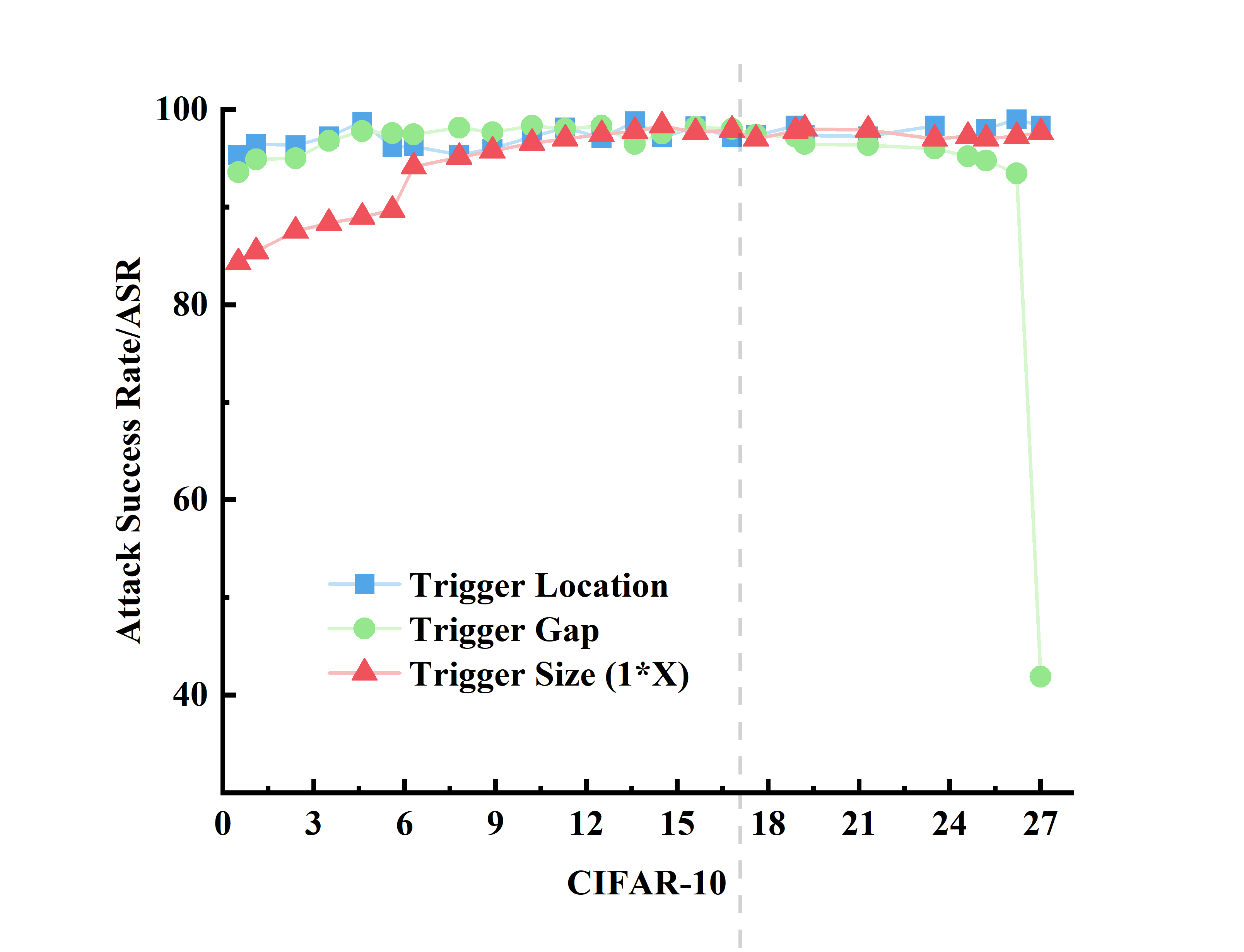}
    \caption{Backdoor Training Framework for SSL.}
    \label{fig:f3}
\end{figure}

As illustrated in Fig~\ref{fig:f3}, experiments on the CIFAR‑10 dataset further validate the conclusions drawn in the main text regarding the influence of trigger mechanisms on the ASR, with results consistent with those observed on STL‑10 and CIFAR‑100. Following the DAB [15] approach, the global trigger was progressively shifted from the top‑left corner of the image to the center and then to the bottom‑right corner (the dashed line indicates that the trigger continues moving along the image boundary after reaching the edge). The results show that changes in trigger location(\( TL\)) have a relatively minor impact on ASR, which can be attributed to the colorful, textured, and structurally diverse nature of CIFAR‑10 images. This high visual complexity and semantic diversity allow the backdoor trigger to blend well across different positions, thereby demonstrating strong robustness to location variations.

In terms of trigger gap(\( TG\)), a significant drop in ASR occurs when four local trigger patterns are placed at the four corners of the image. This is because the local receptive fields of convolutional neural networks struggle to cover widely scattered trigger features, preventing the model from aggregating these dispersed features into a coherent global representation during self‑supervised tasks such as contrastive learning, which ultimately weakens the backdoor effect.

The impact of trigger size(\( TS\)) also aligns with the findings reported in the main text: larger triggers generally lead to higher ASR, but once the size exceeds a certain threshold, the improvement in ASR stabilizes, indicating that further enlargement yields diminishing returns. Overall, ASR is most robust to variations in \( TL\), most sensitive to increases in \( TG\), and exhibits a saturating gain with \( TS\) beyond a critical size. These results further confirm that the trigger‑related properties revealed in the main text generalize across different datasets.

\section{Stability under Different Data Distribution Settings}
 According to our experimental design, we employ Dir(\( \alpha \)) to simulate Non-IID data distributions, where \( \alpha \) varies across 0.05, 0.1, and 10. Table~\ref{tab:adca_performance} presents the final experimental results under Non-IID settings. The results demonstrate that, across all datasets and data distribution setups, ADCA maintains a high ASR, while the drop in ACC compared to the Non-attack scenario is negligible. This indicates that ADCA achieves a strong attack effect while exerting minimal impact on the normal performance of the model. Moreover, as \( \alpha \) increases, the ASR of ADCA improves, further reflecting its stable attack performance under different data distributions.

 \begin{table}[ht]
  \centering
  \begin{tabular}{lcccccc}
    \toprule
    \multirow{2}{*}{Dataset} & \multirow{2}{*}{Setting} & \multicolumn{2}{c}{Non-attack} & \multicolumn{2}{c}{ADCA} \\
    \cline{3-6}
    & & ACC & ASR & ACC & ASR \\
    \midrule
    \multirow{3}{*}{CIFAR-10} & $\alpha$=0.05 & 66.78 & 2.16 & 66.46 & 91.37 \\
    & $\alpha$=0.1 & 71.93 & 5.59 & 70.83 & 94.25 \\
    & $\alpha$=10 & 76.91 & 10.62 & 75.24 & 96.57 \\
    \midrule
    \multirow{3}{*}{STL-10} & $\alpha$=0.05 & 68.48 & 3.62 & 67.63 & 88.14 \\
    & $\alpha$=0.1 & 71.36 & 6.73 & 71.35 & 90.06 \\
    & $\alpha$=10 & 82.64 & 11.34 & 82.78 & 94.93 \\
    \midrule
    \multirow{3}{*}{CIFAR-100} & $\alpha$=0.05 & 59.47 & 3.17 & 58.49 & 90.74 \\
    & $\alpha$=0.1 & 64.17 & 6.36 & 63.52 & 92.51 \\
    & $\alpha$=10 & 71.51 & 12.97 & 71.04 & 95.27 \\
    \bottomrule
  \end{tabular}
\caption{Performance of ADCA under Different Data Distribution Settings.}
\label{tab:adca_performance}
\end{table}

\section{Stability under Different Numbers of Clients}
The number of clients is another critical setup in FL scenarios. To verify the applicability of our proposed method, we evaluate the impact of varying client numbers on ADCA's performance across three datasets. Following previous work~\cite{wu2024towards}, we randomly select 10 clients per round from a total of 25 clients (with 5 being malicious clients) and 20 clients from 80 clients (with 10 being malicious clients) respectively. It is worth noting that at least one malicious client is selected in each round. Additionally, we set up both IID and Non-IID scenarios with \( \alpha \) = 0.1. The experimental results are presented in Table~\ref{tab:client_number_performance1} and~\ref{tab:client_number_performance2}.

\begin{table}[htbp]
  \centering
  \begin{tabular}{lcccccccc}
    \toprule
     \multirow{3}{*}{Dataset} & \multicolumn{4}{c}{10/25 clients}  \\
    \cline{2-5}
    & \multicolumn{2}{c}{IID} & \multicolumn{2}{c}{Non-IID}  \\
    \cline{2-5}
    & ACC & ASR & ACC & ASR \\
    \midrule
    CIFAR-10 & 74.39 & 95.16 & 70.83 & 94.25  \\
    STL-10   & 78.46 & 93.39 & 71.35 & 90.06 \\
    CIFAR-100& 70.03 & 94.62 & 63.52 & 92.51  \\
    \bottomrule
  \end{tabular} 
  \caption{Performance under Different Numbers of Clients(10/25).}
  \label{tab:client_number_performance1}
\end{table}

\begin{table}[htbp]
  \centering
  \begin{tabular}{lcccccccc}
    \toprule
     \multirow{3}{*}{Dataset}  & \multicolumn{4}{c}{20/80 clients} \\
    \cline{2-5}
    & \multicolumn{2}{c}{IID} & \multicolumn{2}{c}{Non-IID} \\
    \cline{2-5}
    & ACC & ASR & ACC & ASR  \\
    \midrule
    CIFAR-10 & 69.31 & 93.37 & 68.19 & 92.74 \\
    STL-10   & 71.44 & 89.36 & 67.25 & 88.32 \\
    CIFAR-100 & 62.96 & 91.63 & 61.76 & 90.68 \\
    \bottomrule
  \end{tabular} 
  \caption{Performance under Different Numbers of Clients(20/80).}
  \label{tab:client_number_performance2}
\end{table}

The results show that on the three datasets CIFAR-10, STL-10, and CIFAR-100, ADCA maintains a high ASR in all settings, with a relatively small decrease in ACC. Specifically, when the number of clients increases from 10/25 to 20/80, regardless of whether the data is IID or Non-IID, the ASR of ADCA decreases slightly but still remains at a high level, and the ACC also shows a corresponding slight downward trend. This indicates that ADCA has good stability under scenarios with different numbers of clients, and can maintain a strong attack effect when the scale of clients changes, while its impact on the normal performance of the model is relatively limited.

\section{Algorithm}
This paper proposes a distributed backdoor attack method (ADCA) for Federated Self-Supervised Learning. In this method, malicious clients decompose the global trigger during the local pre-training phase to seek the optimal decomposition strategy for launching a distributed backdoor attack. Subsequently, these clients collude to form a malicious coalition. They employ an attention mechanism to achieve dynamic parameter aggregation and collaborative learning, thereby exploring the optimal cooperative strategy. The aggregated parameters then serve as the initial state for the next round of training within the coalition, effectively mitigating the dilution of backdoor information caused by benign client updates. The specific algorithm is presented in Algorithm 1.

\begin{algorithm}[htbp]
\caption{Attention-Driven Multi-Party Collusion Attack}
\label{alg:adca}
\begin{algorithmic}[1]
\REQUIRE global malicious client $\tilde{\theta}^t$; local data $\mathcal{D}_k$; malicious coalition $\mathcal{M}$; local trigger $e$, $\varphi^*$;the local model of malicious client $\tilde{\theta}_j$;local epochs $E_{\text{local}}$
\STATE $\tilde{\theta}_j \leftarrow \tilde{\theta}^t$ \hfill $\triangleright$ Initialize from the coalition's model
\FOR{$x \sim \mathcal{D}_k$} 
    \STATE $x_{\text{adv}} \gets R(x, e, \varphi^*)$ \hfill $\triangleright$ Generate backdoor samples
\ENDFOR
\FOR{$E = 1$ to $E_{\text{local}}$}
    \STATE Compute $L_{\text{eff}}$ (trigger recognition)
    \STATE Compute $L_{\text{stealth}}$ (clean representation consistency)
    \STATE $\tilde{\theta}_j \leftarrow \tilde{\theta}_j - \eta \nabla (\lambda_1 L_{\text{eff}} + \lambda_2 L_{\text{stealth}})$ \hfill $\triangleright$ Minimize joint loss
\ENDFOR
\IF{$j \in \mathcal{M}$}
    \STATE $\Delta \theta_j \leftarrow \tilde{\theta}_j - \tilde{\theta}^t$ \hfill $\triangleright$ Compute local model delta
    \STATE Get $\{\Delta \theta_j\}_{j \in \mathcal{M}}$ \hfill $\triangleright$ Collect updates from all malicious clients
    \STATE Extract $S_j$, $C_j$ from each $\Delta \theta_j$
    \STATE $\alpha_j \gets \alpha_j^{\text{spatial}} + \alpha_j^{\text{channel}}$ \hfill $\triangleright$ Attention weights
    \STATE $\tilde{\theta}^{t+1} = \tilde{\theta}^t + \sum_{j=1}^{M} \frac{\alpha_j}{\sum_{k=1}^{M} \alpha_k} \cdot \Delta \theta_j$
    \STATE $\tilde{\theta}_j \gets \tilde{\theta}^{t+1}$ \hfill $\triangleright$ Update with coalition parameters
\ENDIF
\RETURN $\tilde{\theta}_j, \Delta \theta_j$
\end{algorithmic}
\end{algorithm}

\end{document}